\documentclass[lettersize,journal]{IEEEtran}
\usepackage[caption=false,font=normalsize,labelfont=sf,textfont=sf]{subfig}
\usepackage{graphicx}
\usepackage{float}
\usepackage{cprotect}
\usepackage{pgfplots}
\usepackage{boxedminipage}
\usepackage{amsmath}
\usepackage[colorlinks]{hyperref}

\usepackage{cprotect}
\usepackage{amscd,amsmath,amssymb}
\usepackage{amsfonts}
\usepackage{framed}
\usepackage{xspace}
\usepackage{mathtools}
\usepackage{url}
\usepackage{xspace}
\usepackage{verbatim}
\usepackage{verbatimbox}
\usepackage{tcolorbox}
\usepackage{wasysym}
\usepackage{hyphenat}
\usepackage{boxedminipage}
\usepackage[makeroom]{cancel}

\usepackage{tikz}
\usetikzlibrary{trees,arrows,shapes}

\newcommand{\ChunkPreImage}{\textsf{ChunkPreImage}}

\usepackage{fancyvrb}
\usepackage{fvextra}

\usepackage{mdframed}

\def\llncs{0}       

\DeclareMathAlphabet{\mathsl}{OT1}{cmr}{m}{sl}

\ifnum\llncs=0

\newtheorem{thm}{Theorem}[section]
\newtheorem{lem}[thm]{Lemma}
\newtheorem{cor}[thm]{Corollary}
\newtheorem{propo}[thm]{Proposition}
\newtheorem{clm}[thm]{Claim}
\newtheorem{defn}[thm]{Definition}
\newtheorem{assumption}{Assumption}
\newtheorem{rem}[thm]{Remark}
\newtheorem{fct}[thm]{Fact}
\newtheorem{expr}{Experiment}
\newtheorem{cons}[thm]{Construction}
\newtheorem{nte}[thm]{Note}

\newenvironment{theorem}{\begin{thm}\begin{rm}}%
{\end{rm}\end{thm}}
\newenvironment{lemma}{\begin{lem}\begin{rm}}%
{\end{rm}\end{lem}}
\newenvironment{corollary}{\begin{cor}\begin{rm}}%
{\end{rm}\end{cor}}
\newenvironment{proposition}{\begin{propo}\begin{rm}}%
{\end{rm}\end{propo}}
\newenvironment{claim}{\begin{clm}\begin{rm}}%
{\end{rm}\end{clm}}
\newenvironment{remark}{\begin{rem}\begin{em}}%
{\end{em}\end{rem}}
\newenvironment{fact}{\begin{fct}\begin{em}}%
{\end{em}\end{fct}}
\newenvironment{construction}{\begin{cons}\begin{rm}}%
{\end{rm}\end{cons}}
{\end{rm}\end{nte}}

\else

\newtheorem{thm}{Theorem}
\newtheorem{lem}[thm]{Lemma}
\newtheorem{cor}[thm]{Corollary}
\newtheorem{propo}[thm]{Proposition}
\newtheorem{clm}{Claim}
\newtheorem{defn}{Definition}
\newtheorem{assm}{Assumption}
\newtheorem{rem}{Remark}
\newtheorem{fct}[thm]{Fact}

\newtheorem{cons}[thm]{Construction}

{\end{rm}\end{thm}}
{\end{rm}\end{lem}}
{\end{rm}\end{cor}}
{\end{rm}\end{propo}}
{\end{rm}\end{defn}}
{\end{em}\end{assm}}
{\end{rm}\end{clm}}
{\end{em}\end{rem}}
{\end{em}\end{fct}}
{\end{rm}\end{cons}}

\fi

\ifnum\llncs=0

\newlength{\saveparindent}
\newlength{\saveparskip}
\newcount\proofqeded
\newcount\proofended

\def\qed{{\hspace{2pt}\rule[-1pt]{3pt}{9pt}}
\end{rm}\addtolength{\parskip}{-0pt}
\setlength{\parindent}{\saveparindent}
\global\advance\proofqeded by 1 }

 {\ifnum\proofqeded=\proofended\qed\fi \global\advance\proofended by 1
  \medskip}
\makeatletter
\def\proofstart{\@ifnextchar[{\@oprf}{\@nprf}}
\def\proofsketchstart{\@ifnextchar[{\@osprf}{\@nsprf}}
\def\@oprf[#1]{\begin{rm}\protect\vspace{6pt}\noindent{\bf Proof of #1:\ }%
\addtolength{\parskip}{5pt}\setlength{\parindent}{0pt}}
\def\@osprf[#1]{\begin{rm}\protect\vspace{6pt}\noindent{\bf Sketch of
Proof of #1:\ }
\addtolength{\parskip}{5pt}\setlength{\parindent}{0pt}}
\def\@nprf{\begin{rm}\protect\vspace{6pt}\noindent{\bf Proof:\ }%
\addtolength{\parskip}{5pt}\setlength{\parindent}{0pt}}
\def\@nsprf{\begin{rm}\protect\vspace{6pt}\noindent{\bf Proof Sketch:\ }%
\addtolength{\parskip}{5pt}\setlength{\parindent}{0pt}}

\fi

\ifnum\llncs=1

\newcommand{\headingb}[1]{\vspace{-5pt}\subsubsection{#1}}
\else

\newcommand{\headingb}[1]{
}
\fi

\newcommand{\etal}{{\em et al.}}

\newcommand{\remove}[1]{}

\ifnum\llncs=0

\makeatletter
\def\appearsin#1{\gdef\@appearsin{#1}}

\def\maketitle{\par
 \begingroup
 \def\thefootnote{\fnsymbol{footnote}}
 \def\@makefnmark{\hbox
 to 0pt{$^{\@thefnmark}$\hss}}
 \if@twocolumn
 \twocolumn[\@maketitle]
 \else \newpage
 \global\@topnum\z@ \@maketitle \fi\thispagestyle{plain}\@thanks
 \endgroup
 \setcounter{footnote}{0}
 \let\maketitle\relax
 \let\@maketitle\relax
 \gdef\@thanks{}\gdef\@author{}\gdef\@title{}\gdef\@appearsin{}
          \let\thanks\relax}
\def\@maketitle{\newpage
 \vskip 0.5in \begin{center}
 {\LARGE \@title \par} \vskip 1.5em {\large \lineskip .5em
\begin{tabular}[t]{c}\@author
 \end{tabular}\par}
 \vskip 1em {\normalsize \@date} \end{center}
 \par
 \vskip 1.5em}

\fi

\newcommand{\ugly}{{\color{red}\times}}
\newcommand{\sexy}{{\color{green}\surd}}
\newcommand{\munnezza}{\ugly}

\definecolor{njmGray}{gray}{0.8}

\newcommand{\Verify}{{\sf Verify}}

\newcommand{\trans}{\sf tr}
\newcommand{\miner}{\sf Miner}

\newcommand{\Prove}{\mathsf{Prove}}

\newcommand{\SHARound}{{\sf SHARound}}

\newcommand{\nota}[1]{{\bf Nota: } #1 \par}
\renewcommand{\nota}[1]{}

\newcommand{\NP}{{\sf NP}}

\newcommand{\PreImage}{\mathsf{PreImage}}

\renewcommand{\nota}[1]{}

\newcounter{itemcount}

\newcommand{\newreptheorem}[2]{
\newenvironment{rep#1}[1]{
 \def\rep@title{#2 \ref{##1}}
 \begin{rep@theorem}}
 {\end{rep@theorem}}}
\makeatother

\newcommand{\inte}{intervals}

\begin{document}

\title{Towards Data Redaction in Bitcoin\thanks{This result appeared in IEEE Transactions on Network and Service Management Journal\cite{OURS}. V. Botta and I. Visconti are with the University of Salerno, Fisciano, Italy (email: botta.vin@gmail.com, visconti@unisa.it). V. Iovino is with the Aragon Association, Zurich, Switzerland (email: viovino@unisa.it).}
}
	\date{}

\author{Vincenzo~Botta,
        Vincenzo~Iovino,
        and~Ivan~Visconti
}

\maketitle

\begin{abstract}
A major issue for many applications of blockchain technology is the tension between immutability and compliance to regulations. For instance, the GDPR in the EU requires to guarantee, under some circumstances, the right to be forgotten. This could imply that at some point one might be forced to delete some data from a locally stored blockchain, therefore irreparably hurting the security and transparency of such decentralized platforms. 

Motivated by such data protection and consistency issues, in this work
we design and implement a mechanism for securely {\em deleting} data from Bitcoin  blockchain.
We use zero-knowledge proofs to allow any node to delete some data from Bitcoin transactions, still preserving the public verifiability of the correctness of the spent and spendable coins. 
Moreover, we specifically use STARK proofs to exploit the transparency that they provide.

Our solution, unlike previous approaches, avoids the complications of asking nodes to reach consensus on the content to delete. In particular, our design allows every node to delete some specific data without coordinating this decision with others. In our implementation, data removal can be performed (resp., verified) in minutes (resp., seconds) on a standard laptop rather than in days as required in previous designs based on consensus.
\end{abstract}

	
	\section{Introduction}\label{sec:intro}
\paragraph{Data protection
and Bitcoin blockchain}

Bitcoin blockchain~\cite{Nakamoto09} is sometimes described as a censorship-free financial platform due to the inability of governments and institutions of blocking and restricting the creation and transfer of bitcoins.
Recent discoveries (e.g.,~\cite{FC:MHHZMHW18,BisMerSan19}) raise concerns on the immutability of Bitcoin blockchain.
Indeed, Bitcoin blockchain can also include non-financial data stored in transactions as proven by Matzett \etal~\cite{FC:MHHZMHW18}, that discovered some contents related to child pornography and to dark web services. This 
motivates the problem that at some point in some countries it could be  illegal  to store the blockchain, which is however a fundamental requirement for the transparency of Bitcoin. This problem was also more recently discussed in~\cite{Sch21}.

In addition, the European data protection regulation, the
GDPR~\cite{EU16}, enforces the ``right to be forgotten'', according to which individuals have the right to ask for deletion of 
their personal data if certain conditions apply. This regulation seems to clash with the immutability properties of so-called permissionless blockchains like Bitcoin.

\paragraph{Common ways to encode arbitrary data on the blockchain} 
In the following, we assume the reader to be familiar with the format of Bitcoin transactions and,  for concreteness, we 
will only describe the mechanics of Bitcoin that are useful for our work. Informally, 
a script is a list of instructions including operation codes. A Bitcoin transaction is useful to transfer money specifying:
 a unique identifier \verb|TXID|, a list of input scripts called \verb|scriptSig|, a list of output scripts called \verb|scriptPubKey|, and a value \verb|VOUT|. A special transaction known as  Coinbase generates money. 

Each \verb|scriptSig| in a transaction consumes the output of a previous transaction that is then locked in some \verb|scriptPubKey| scripts. 
Each input and output script can be written in Bitcoin using a specific scripting language called Script,
that is  a stack-based language intentionally not Turing-complete (e.g., no loops). 
A transaction is valid
if concatenating the \verb|scriptSig| script and the \verb|scriptPubKey| script, the resulting script 
is evaluated successfully (i.e., during the evaluation nothing triggers a failure and in the end the top of the stack corresponds to True). For details  see~\cite{script}. 
\verb|OP_RETURN| is an  operation code that when executed
ends unsuccessfully the execution.
Each time a valid transaction is executed, coins associated to the output script \verb|scriptPubKey| are spent and sent to the owner of the input script \verb|scriptSig| of the valid transaction. The amount of spendable digital currency is stored in the unspent transaction output (UTXO) database.

We describe now the two most common and natural ways to encode arbitrary data in Bitcoin transactions.

\begin{itemize}
	\item{Coinbase transactions.} 
	A coinbase transaction is a transaction in which the field  \verb|scriptSig| can  contain arbitrary data.
	For instance, the \verb|scriptSig| field of the genesis coinbase transaction, identified by \verb|TXID| $4a5e1e4baab89f3a32518a88c31bc87f618f76673e2cc \allowbreak 77ab2127b7afdeda33b$, is (decoded as) the string ``The Times 03/Jan/2009 Chancellor on brink of second bailout for banks''.
	
	\item{Data output transactions.} The \verb|OP_RETURN| mechanism can be used in the following {\em general form}: \verb|... OP_RETURN <DATA> ...|, where \verb|<DATA>| is a string of at most 83 bytes\footnote{This type of transaction also includes a field needed to deal with strings of variable length. For simplicity we omit 
	this field from our analysis.}.
	The \verb|OP_RETURN| functionality was actually introduced in Bitcoin with the purpose of
	allowing to store data on the blockchain. 
	
\end{itemize}

\subsection{Previous Solutions and Their Limitations}\label{sec:ourscenario}
A first approach to data redaction in  blockchains was proposed by Ateniese \etal~\cite{EUROSP:AMVE17} that mainly tackled the permissioned setting and thus remained ineffective for Bitcoin. 
In~\cite{EPRINT:PudDmiCap17,SP:DeuMagThy19,Reparo20} the problem of redacting a permissionless blockchain is solved using consensus protocols. Each of these works presents different voting procedures that users have to perform to decide if a proposed modification can be stored on the blockchain or should be refused.

In Section~\ref{sec:related} we will provide a more extensive survey of related works and we will compare them to our solution.

\paragraph{Our main questions}
In light of the above discussion, we have the following natural open questions:
	 {\em Can data be in general deleted by individual nodes (i.e., also without the use of voting protocols among nodes) preserving public verifiability? 
	 Can the update be realized without requiring a {\em hard} fork? 
	 If not in general, 
	 in which restricted cases is it instead
	 possible (if any)?}
	The obvious requirement is that data redaction should  not hurt the public verifiability of the correct state of the blockchain (e.g., it should not affect the reliability of the UTXO database).

\subsection{Our Scenario and Results}\label{subsec:res}
In our work we envision a scenario in which a Bitcoin node storing the full blockchain wants to delete some data encoded either in the coinbase or in a data output transaction. Our approach completely deviates  from previous ones in that it relies on  
individual redaction rather than on jointly decided redaction.
Each individual redaction will not be replicated by the Bitcoin network. Indeed, our solution permits a scenario in which a set of nodes can delete some content due to imposition by an authority whereas other nodes can still keep such content. We remark that our solution  guarantees that the Bitcoin financial state (i.e., the UTXO database)  remains unchanged.

Differently from~\cite{EPRINT:PudDmiCap17,SP:DeuMagThy19,Reparo20} that achieve redactions using voting protocols, we do not seek for consensus on redaction. On the other hand, we do require {\em public verifiability}, so that
correctness of transactions in a redacted blockchain can still be verified.  We remark that  by public verifiability we mean the ability of verifying the consistency of Bitcoin financial state (i.e., the correctness of UTXO), while instead in Thyagarajan \etal~\cite{Reparo20} they refer to the accountability of redactions.
In particular, our solution guarantees the ability of verifying that the chain of blocks is consistent; the verification will not be executed by just using the hash function as in the standard Bitcoin protocol but, as we will see later, by also verifying zero-knowledge proofs.
Furthermore, the public verifiability is transparent in the sense that the verification will not be based on parameters that depend on secrets owned by trusted parties.

\paragraph{Our results}
We summarize our results as follows.
\begin{itemize}
	\item We answer the above questions by carefully analyzing Bitcoin protocol and showing in which cases data redaction may be harmful and why it is not possible {\em in general} to delete content from Bitcoin.
	\item Then, we provide a solution to sanitize (i.e., safely allowing to remove data from the blockchain) Bitcoin  in those well known cases where arbitrary data can be encoded in transactions; we show how to tweak the Bitcoin blockchain to enable such data redaction mechanism.
	\item We show that our data redaction mechanism for Bitcoin is practical. We present our implementation and show how to sanitize Bitcoin with concrete examples.
\end{itemize}
We stress that the types of redaction we consider only regard auxiliary data  that can be inserted in coinbase or data output transactions that when removed do not change the impact of the transaction w.r.t. the UTXO database. Moreover, our method does not change the chaining performed by linking heads of blocks (including roots of Merkle trees of transactions) and as such it has no impact on attackers attempting to perform double-spending attacks.
\paragraph{Extending our results to other permissionless blockchains}
It is natural to ask whether our techniques could be also fruitful for other permissionless blockchains.
In our solution we focus on the specific mechanism used in Bitcoin to add illicit content in the Bitcoin blockchain (i.e., \verb|OP_RETURN| and coinbase transactions). Our solution is specific to these mechanisms that permit to keep separate the UTXO database from data to be removed. In order to apply our technique to a different blockchain one needs to figure out how illicit data can be encoded in this  blockchain, and if those mechanisms permit to separate the potentially illegal contents to be removed from the actual state of the blockchain (e.g., the equivalent of the UTXO database).
Moreover, even in case it is possible to somehow apply our technique, it is extremely relevant to check if the update would cause a hard or a soft fork in the blockchain. Therefore, a successful application of our technique strongly depends on the inner details of the target blockchain.
Another point to take into account is that in Bitcoin a new block is created each 10 minutes. Therefore, it is possible to exploit this time window to redact on the fly some transactions containing the \verb|OP_RETURN| mechanism. In this scenario, once a node receive a data output transaction $t$, if $t$ contains some illicit content (e.g., $t$ contains some specific keywords), the node can decide to redact directly $t$ without writing the illicit content in the storage. 

\subsection{Is Generic Data Redaction Possible in Bitcoin?}
One might think that deleting data from data output transactions is innocuous since strings following \verb|OP_RETURN <DATA>| have no impact
on the UTXO database.
However, there is an important issue: at the bootstrap time each node downloading the Bitcoin blockchain should check the consistency of blocks and transactions.
If a new Bitcoin participant $P$ downloads the entire blockchain from a full node $N$, following the rules of Bitcoin, $P$ will check the chain consistency computing the following steps:
\begin{enumerate}
	\item $P$ hashes all the transactions contained in each block;
	\item $P$ uses these hashes as leaves of a Merkle tree and computes the Merkle root of the tree;
	\item $P$ verifies that the Merkle root obtained from the transactions in each block is equal to the Merkle root stored in the block header.
\end{enumerate}
If a transaction has been modified then $P$ would notice the corruption of the blockchain. Obviously requiring $P$ to just trust the blockchain provided by $N$, despite a failure in the check, is not acceptable.
To solve this bootstrap verification issue, we propose the following solution (the description is simplified): every time a full node $N$ has to delete data from a transaction $t$ in a block stored on the blockchain, $N$ executes the following steps.
\begin{enumerate}
	\item $N$ modifies $t$ in a new transaction $t'$ where all data to be redacted in $t$ are substituted with zeroes. Moreover $N$ stores the hash $h$ of $t$. 
	\item $N$ replaces $t$ with $t'$ in the locally stored blockchain, without recomputing the root of the Merkle tree, leaving $h$ in the leaf.
	\item $N$ generates a non-interactive zero-knowledge proof (NIZK)\footnote{Here we are presenting the a generic construction that can use any NIZK, but, for efficiency reasons, we  substitute NIZKs with STARKs later in the presentation. For the definition of NIZKs and STARKS see Section~\ref{sec:zk}.} of the following statement: there exists a sequence of bytes that substituted in $t'$ in specific harmless positions would produce a transaction $t''$ such that the hash of $t''$ is $h$. 
	\item Every time someone requests the blockchain to $N$, $N$ sends the blockchain containing each modified $t'$ together with the NIZK proofs and the statements.
\end{enumerate}

When $P$ downloads the blockchain from $N$, $P$ receives the modified transactions with the NIZK proofs and the statements for each block containing modified transactions. $P$ then executes the following steps.
For each block $B$, and 
for each redacted transaction in $B$, $P$ runs the verifier of the NIZK. If the NIZK proof is not valid then $P$ marks $B$ as invalid. If the NIZK proof is valid or if there is no redacted transaction then $P$ computes a Merkle tree as follows. For each non redacted transaction $P$ puts in the leaf of the Merkle tree the hash of the transaction; for each redacted transaction, $P$ extracts the hash $h$ from the statement of the NIZK proof and uses this hash as leaf of the Merkle tree.
$P$ computes the root of the Merkle tree and considers $B$ a valid block only if the computed Merkle root is equal to the Merkle root contained in the header of $B$ downloaded from $N$.

If all checks are successful, then $P$ assumes that the downloaded blockchain is correct.
Notice that the consistency of the chain is guaranteed also by the NIZK proofs. Such proofs ensure consistency when a redaction is done for a transaction belonging to a block $B$, still leaving the Merkle root in the header of $B$ unchanged. The only inconsistency that holds is between the old hash and the hash of the new redacted transaction but this inconsistency is fixed by the NIZK proof.

\section{Related Work and Comparison}\label{sec:related}
Ateniese \etal~\cite{EUROSP:AMVE17} proposed the first protocol for illicit content deletion from blockchains. Their solution is simple and efficient but, unfortunately, mainly targets the permissioned setting and cannot be adapted to Bitcoin. Unlike ours, in their approach a deletion does not leave trace and goes unnoticed to users not participating in the redaction. 
The solution of Ateniese \etal\ is based on the concept of chameleon hash function, essentially hash functions endowed with trapdoors that allow to find different preimages to a given hashed value. The drawback of solutions based on such kind of cryptographic tool is that the trapdoor should be kept secret or shared among a set of authorities. In our solution instead we do not assume any set of authorities to share some secrets needed for the redaction.

Puddu \etal~\cite{EPRINT:PudDmiCap17} provided a more complex protocol for dealing with redactions of harmful content. They proposed a protocol in which users can set alternate versions, called ``mutations'', of their transactions that can be later activated after running an expensive MPC protocol. A request of a modification has to be approved by means of a voting procedure based on proofs of work. In their solution, only the creator of a transaction can allow modifications, thus preventing deletion of content inserted by malicious parties. The main drawback of their solution is that the ability of ``mutating some content''  has to be explicitly set by the miners and so malicious miners can simply bypass the mutation mechanism. Moreover, mutation of some content has a cascade effect on any subsequent transaction, thus incurring a huge performance penalty.

Deuber \etal~\cite{SP:DeuMagThy19} proposed a novel redactable blockchain protocol that can be integrated in Bitcoin. In their protocol, each user can propose a modification by writing the proposal on the blockchain. The redaction proposal is subject to a voting procedure based on consensus and computational power. The Deuber \etal's proposal 
requires a voting procedure performed online on the blockchain 
whereas in our protocol each node can {\em individually} perform a deletion without the need of interaction with other nodes. 
Deuber \etal's introduce ``public verifiability'' that in their case consists of the ability of tracing redactions. In our protocol redactions can be traced as well.

Thyagarajan \etal~\cite{Reparo20} proposed Reparo, a protocol that improves Deuber \etal's solution with the property of ``Reparaibility of Existing Content'' (REC), that is the possibility of redacting or modifying blocks that are inserted in the blockchain before the software update that includes the redaction protocol is performed. 
As in Deuber \etal, Reparo is based on expensive and interactive consensus protocols that requires several {\em days} to be run as opposed to our protocol in which deletion can be performed in few {\em minutes}. Both Thyagarajan \etal\ and Deuber \etal\ do not guarantee individual deletion, meaning that it is not possible for a single node to delete data locally without starting the voting procedure.

Florian \etal~\cite{FHBS19} proposed a different approach in which  nodes do not completely validate the chain and have to trust others, while in our solution a blockchain subject to data redaction can be completely validated by each node.

Grigoriev \etal~\cite{GS20} proposed a data redaction mechanism based on the RSA cryptosystem. Their work focuses on permissioned blockchains. In their construction each block $B_i$ of the blockchain can be seen as a triplet $(P_i, C_i, X_i)$, where $P_i$ is the immutable prefix, $C_i$ is the actual content and $X_i$ is a suffix. When a block $B_i$ must be redacted, a central authority $H$ should have a key that allows $H$ to change the content $C_i$ of $B_i$ with a new content $C_i'$, selecting a new suitable $X_i'$.

Dousti \etal~\cite{DK20} proposed four attacks against  redactable blockchain solutions. The first attack is specific to the protocol of Grigoriev \etal: the attacker can craft two new blocks $B$ and $B'$, append $B$ to the ledger and at any point in time the adversary can change $B$ with $B'$ without involving the administrator. The second attack applies both to the Ateniese \etal\ and Grigoriev \etal's protocols and states that an adversary can always change a redacted block of the blockchain with the original block. These first two attacks are only applicable to the permissioned redactable blockchains. 
The third and forth attacks introduced by Dousti \etal\ are specific for redaction techniques based on votes.
The third attack considers an attacker who erases blocks containing votes for a chosen redaction. The fourth attack considers a scenario in which the adversary controls the 49\% of the miners so that the votes are strongly influenced by the adversary. We will show a more sophisticated version of this last attack 
in Section~\ref{subsec:quality}.

\subsection{Quality of the Redaction Decision}\label{subsec:quality}
Another issue in the protocols of Deuber \etal\ and Thyagarajan \etal, is that by instantiating those protocols for Bitcoin, even if the adversary does not have half of the global hash power, it can still control the voting procedure.
According to Garay \etal~\cite{EC:GarKiaLeo15}, an adversary controlling a fraction $t$ of the hash power can control up to a fraction $\frac{t}{1-t}$ of the blocks in the chain.
Thyagarajan \etal\ (see~\cite[Appendix E]{Reparo20}) and Deuber \etal\ (see~\cite[Section 5.2]{SP:DeuMagThy19}) concretely suggest to consider a redaction in Bitcoin accepted if it received more than 50\% (i.e., 
$\frac{1}{2}+\delta$, for any  $\delta>0$) of the votes in the $1024$ blocks after the redaction proposal. We call ``voting threshold'' the threshold of votes needed for a redaction to be accepted in the protocol.
Due to Garay \etal's analysis, we observe that a voting threshold parameter of $\frac{1}{2}+\delta$ is too optimistic since it can allow
an attacker owning $\frac{1+2\delta}{3+2\delta}$ of the hash power to control the voting procedure (and thus the ability of redacting the blockchain). Indeed, if the adversary controls  $\frac{1+2\delta}{3+2\delta}$ of the hash power, then the adversary can control $\frac{t}{1-t}=\frac{(1+2\delta)/(3+2\delta)}{1-(1+2\delta)/(3+2\delta)}=\frac{1+2\delta}{2}=\frac{1}{2}+\delta$
of the blocks in the chain, thus obtaining the majority of the votes.

We say that a redaction
mechanism in Bitcoin achieves {\em $t$-quality} if no adversary controlling a fraction $t<\frac{1}{2}$ 
of the resources  (i.e., the hash power in proof-of-work blockchains) can succeed
in the attack aiming at forcing data redaction  
when all other nodes are against redaction. As argued above, the redaction mechanisms in the aforementioned works are such that whatever voting threshold $>\frac{1}{2}$ 
is selected there exists a value 
$t<\frac{1}{2}$ 
such that an adversary controlling a fraction $t$ of the resources 
succeeds in the attack.

For instance, Reparo instantiated with voting threshold $\frac{1}{2}+\delta$, for $\delta>0$  (as suggested by the authors), and assuming adversaries owning $t=\frac{1+2\delta}{3+2\delta}$ of the total hash power does not satisfy $\frac{1+2\delta}{3+2\delta}$-quality. Indeed, following what stated by Garay \etal\, as previously shown, it holds that  $\frac{t}{1-t}=\frac{1}{2}+\delta$, meaning that the adversary can control enough blocks to redact the content on other nodes even if the majority is against the redaction. In general, let $f$ be the voting threshold, if the hash power of the adversary is at least  $t=\frac{f}{1+f}$, the adversary can always succeed in the attack (indeed for each $f\leq 1-\delta$, for $\delta>0$, $t$ is $\frac{1-\delta}{2-\delta}<\frac{1}{2}$).

Moreover, Reparo does not satisfy  $\frac{2}{5}$-quality when instantiated with voting threshold $\frac{2}{3}$ and the adversary controls a fraction $t=\frac{2}{5}$ of the hash power, indeed $\frac{t}{1-t}=\frac{2/5}{1-2/5}=\frac{2}{3}$. However, Reparo for voting threshold parameter $\frac{3}{4}$ 
satisfies $\frac{2}{5}$-quality since
assuming that and adversary controls $t=\frac{2}{5}$ of the total hash power
we have that
$\frac{t}{1-t}=\frac{2/5}{1-2/5}=\frac{2}{3}<\frac{3}{4}$ (i.e., the adversary does not control enough blocks to force the redaction of contents if all other nodes are against redaction).

A natural goal is that whatever threshold $f$ of voters is used to reach consensus on redaction, no adversary controlling less than half of the hash power should be able to perform a redaction.
Notice that redaction mechanisms based on voting are also prone to bribing attacks in which a player can bribe others (paying out-of-band or with cryptocurrency) towards controlling what should be redacted and what should not.

In Tables~\ref{tbl:comparison} and~\ref{tbl:comparisonthree} we compare  known results with ours.

\begin{table}
\caption{Comparison of our solution with the state of the art in redaction of blockchains. \label{tbl:comparison}}

\centering
\resizebox{\columnwidth}{!}{
\begin{tabular}{|c|c|c|c|}
   \hline
	$\substack{\textrm{Solution}}$&$\substack{\textrm{Permissionless}}$&$\substack{\textrm{Publicly Verifable Deletion}}$&$\substack{\textrm{REC}}$\\
   \hline
	$\substack{\textrm{Ateniese \etal\ \cite{EUROSP:AMVE17}}}$&$\substack{\munnezza}$&$\substack{\munnezza}$&$\substack{\munnezza}$\\
   \hline
	$\substack{\textrm{Puddu \etal\ \cite{EPRINT:PudDmiCap17}}}$&$\substack{\sexy}$&$\substack{\ugly}$&$\substack{\ugly}$\\
   \hline
	$\substack{\textrm{Deuber \etal\ \cite{SP:DeuMagThy19}}}$&$\substack{\sexy}$&$\substack{\sexy}$&$\substack{\ugly}$\\
	\hline
	$\substack{\textrm{Thyagarajan \etal\ \cite{Reparo20}}}$&$\substack{\sexy}$&$\substack{\sexy}$&$\substack{\sexy}$\\
	\hline
	$\substack{\textrm{This work}}$&$\substack{\sexy}$&$\substack{\sexy}$&$\substack{\sexy}$\\
\hline
 \end{tabular}
 }

\end{table}

\begin{table}
\caption{Comparison of our solution with the protocols of Deuber \etal\ and Thyagarajanan \etal\ when instantiated for Bitcoin. For The individual deletion and t-quality properties see Section~\ref{subsec:res} and Section~\ref{subsec:quality}.\label{tbl:comparisonthree}}
\centering
\resizebox{\columnwidth}{!}{
\begin{tabular}{|c|c|c|}
   \hline
	$\substack{\textrm{Solution}}$& \shortstack{$\substack{\textrm{Individual Deletion}}$\\ $\substack{\textrm{(no consensus)}}$}&$\substack{\textrm{\shortstack{$t$-Quality}}}$\\
   \hline
	$\substack{\textrm{Deuber \etal\ \cite{SP:DeuMagThy19}}}$&$\ugly$&$\substack{\textrm{Failure for any }t\geq\frac{1+2\delta}{3+2\delta}}$\\
	\hline
	$\substack{\textrm{Thyagarajan \etal\ \cite{Reparo20}}}$&$\ugly$&$\substack{\textrm{Failure for any }t\geq\frac{1+2\delta}{3+2\delta}}$\\
	\hline
	$\substack{\textrm{This work}}$&$\sexy$& $\substack{\textrm{Success for any } t<\frac{1}{2}}$\\
\hline
 \end{tabular}
 }
\end{table}

	\section{Preliminaries}
\subsection{Bitcoin in a Nutshell}
Bitcoin~\cite{Nakamoto09} is a permissionless blockchain system that allows users to perform electronic payments without the need of a trusted party.
In Bitcoin there are two specific standard transactions called respectively data output transaction and coinbase transaction 
allowing to store arbitrary data on the Bitcoin blockchain. The data output transaction was added to publish arbitrary data using a provably unspendable \verb|scriptPubKey| script in which the specific opcode \verb|OP_RETURN| is used. The coinbase transaction is a specific transaction used in Bitcoin as first transaction of a new block of the blockchain. The \verb|scriptSig| of a coinbase transaction can be used to store any arbitrary data since the coinbase transaction is used to generate new coins without redeeming money coming from previous transactions.

In Bitcoin all transactions are public and can be viewed and checked by everyone.
Bitcoin is based on proofs-of-work: every time a miner $\miner$ wants to publish a new block, $\miner$ needs to solve a cryptographic puzzle that consists of finding a value whose hash has a certain number of zero leading bits. 

The following operations are performed in Bitcoin network:
\begin{itemize}
    \item every time a party generates a new transaction $\trans$, $\trans$ is sent to all nodes;
    \item each miner collects new transactions into a block;
    \item each miner works on finding generating a proof-of-work (i.e., a solution to the cryptographic puzzle) for its block;
    \item when a miner gets a proof-of-work, it broadcasts the block including the proof-of-work to all nodes;
    \item nodes accept the block only if all transactions in it are valid and the proof-of-work is correct;
    \item miners express their acceptance of the block by working on creating the next block in the chain, using the hash of the accepted block.
\end{itemize}
Miners consider the longest work-weighted chain to be the correct one and will keep working on extending it.

In order to save disk space, it is possible to delete locally some old transactions of Bitcoin maintaining the UTXO database. It is possible to delete local data without breaking the block's hash, since transactions are hashed in a Merkle Tree with only the root included in the block's hash.
A full network node is a node that maintains the entire Bitcoin history.

\subsection{Bitcoin Scripts}\label{subsubsec:script}
Bitcoin uses a scripting language to express conditions to transfer coins. Each transaction $t$, generated by a party $P$, contains a set of input scripts. Each input script, called
 \verb|scriptSig|, is used to redeem coins from a previous transaction $t'$. An input script is a witness proving that $P$ can spend coins allocated in $t'$. $P$ can redistribute those coins writing in $t$ a set of output scripts, where each of these output scripts is called  \verb|scriptPubKey|. If a party $P'$ can generate a transaction $t''$ that contains an input script that is a witness for one of the output scripts in $t$, then $P'$ can spend the coins associated to the output script of $t$.

Bitcoin scripting language includes opcodes, that are instructions for the scripts. In this work we will focus on two main opcodes, \verb|OP_CHECKSIG| and \verb|OP_RETURN|.
The \verb|OP_CHECKSIG| is used in Bitcoin  to verify a signature taking
as input a public key and a signature. Such opcode outputs True if the signature passes the check and False otherwise.
We now recall the steps of a Bitcoin node to obtain the message on which the signature verification is performed.
Let $t_1$ be a previous transaction and $t_2$ be a new transaction that wants to redeem the output script $o_1$ of $t_1$ using input script $i_2$. The output script $o_1$ includes both  \verb|OP_CHECKSIG| and a public key. The user that generates $t_2$ is supposed to have the corresponding secret key. Indeed, the input script $i_2$ must include a signature computed with such secret key. The message on which the signature should be verified is computed as follows\footnote{We skip the description of the steps that are not relevant for our work.}.
From $o_1$ that includes the \verb|OP_CHECKSIG|, a new script $o_1'$ is created. The script $o_1'$ consists of data from the most recently parsed \verb|OP_CODESEPARATOR| until the end of $o_1$.
All remaining \verb|OP_CODESEPARATOR|s are removed from $o_1'$.
Let $t_{2copy}$ be a clone of $t_2$.
Each input script in $t_{2copy}$ is set to the empty string.
Finally, $i_2$ in  $t_{2copy}$ is set to $o_1'$.

The node that wants to verify the signature hashes twice  $t_{2copy}$ with SHA256. The resulting string is the message on which the signature is verified.

The opcode \verb|OP_RETURN| is used to publish the standard locking script \verb|NULLDATA|, also called data output transaction, since it is provably unspendable and discarded from storage in the UTXO database\cprotect\footnote{As mentioned in Section~\ref{sec:intro}, it is possible to insert an \verb|OP_RETURN| opcode in a spendable transaction. See \url{https://learnmeabitcoin.com/technical/nulldata} (accessed 2022/10/04).}. The opcode \verb|OP_RETURN| has two parameters, the first parameter is the number of bytes to store in the transaction and the second parameter consists of the ``free'' bytes to store in the script.

\subsection{NIZK, SNARKs and STARKs}\label{sec:zk}
Let $R$ be an efficiently computable binary relation.
For pairs $(x, w) \in R$ we call $x$ the statement and $w$ the witness.

A non-interactive zero-knowledge (NIZK) argument system for a relation $R$ consists of the following pair of probabilistic polynomial-time (PPT) algorithms (with implicit access to a random oracle (RO) $O$) that must satisfy properties called completeness, soundness and zero knowledge\footnote{We defer to~\cite{INDOCRYPT:FKMV12,PKC:BerFisWar15} for formal definitions.
} that we informally report below:
\begin{itemize}
	\item $\Prove(x,w)$: this is a PPT algorithm that takes as input
a statement $x$ and a witness $w$ for $R$, and with oracle access to $O$ produces a proof $\pi$.
\item $\Verify(x,\pi)$: this is a deterministic polynomial-time algorithm that takes as input a statement $x$ and a proof $\pi$, and with oracle access to $O$ outputs $1$ if the proof is accepted and $0$ otherwise.
\end{itemize}

\begin{itemize}
	\item {\sf Completeness:} An honest prover convinces an honest verifier with overwhelming probability.
	\item {\sf Soundness:} The probability that a PPT dishonest prover convinces an honest verifier on a false statement is negligible.
	
\item {\sf Zero knowledge:} 
The proof computed by $\Prove$ does not reveal any additional information.

\end{itemize}

If a pair of PPT algorithms $(\Prove,\Verify)$ satisfies all previous properties except for the ZK property we say that $(\Prove,\Verify)$ is an argument system.

Moreover, we say that a NIZK is an argument of knowledge (NIZKAoK) if it satisfies 
the following property:
\begin{itemize}
    \item {\sf Extractability:} Given a malicious PPT prover, there exists an efficient extractor algorithm $E$ such that if the prover produces with non-negligible probability an accepting proof $\pi$ for a statement $x$, then $E$ with access to the prover outputs a witness $w$ for $x$. 
\end{itemize}

Zero-knowledge succinct non-interactive arguments of knowledge (zk-SNARKs) are  NIZKAoK such that the proof has short size and the verification of the proof is fast.

We call prover the user that runs the $\Prove$ algorithm, while the user that runs the $\Verify$ algorithm is the verifier. zk-SNARKs in some cases require parameters that are generated by a trusted party that carefully deletes any auxiliary information. In our work we use zk-SNARKs that are transparent, meaning that no trusted parameter is used.
Such systems are called zero-knowledge succinct transparent argument of knowledge (STARKs)~\cite{C:BBHR19,TCC:BenChiSpo16}.

In a SNARK/STARK the algorithms must satisfy completeness, soundness and extractability,
and moreover the following succinctness property:
the verifier must run in polynomial time in $\lambda$ plus the size of the statement $x$; moreover, the proofs scale sublinearly in the size of the witness for  $x$.

As already specified, in our approach we will use zk-STARKs but we stress here that succinctness with respect to the size of the claim is not crucial for our redaction mechanism. In the following we refer to zk-SNARKs/zk-STARKs when we talk about SNARKs/STARKs.

We defer the reader to~\cite{C:FiaSha86,Micali00,C:Kilian95,EC:GGPR13,SP:BBBPWM18,C:BBHR19,TCC:BenChiSpo16} for detailed definitions of NIZKs, SNARKs and STARKs.

	\section{Our Bitcoin Sanitizer}\label{sec:sanitizer}
We first show that the problem of data redaction from Bitcoin boils down to computing and verifying  NIZK proofs for a conceptually simple (class of) statements. Then we will show how to implement proofs for such statements in an {\em efficient} way by losing only harmless information. 

\subsection{The General Statement}
Let $h$ and $X_1,\ldots,X_n,X_{n+1},n\ge 1$ be public strings (possibly empty) and let $H$ be the SHA256 function used in Bitcoin.
Consider the following statement $\PreImage_{h,X_1,\ldots,X_{n+1}}$:
$$\exists y_1,\ldots,y_n:\ H(X_1||y_1\cdots X_n||y_n||X_{n+1})=h,$$ in which we implicitly assume that the indices of the substrings $y_1,\ldots,y_n$ and their lengths are public and part of the statement and appear only in {\em allowed} positions, that is after an \verb|OP_RETURN| opcode or in a coinbase transaction \verb|scriptSig| field (i.e.,
in positions where deletion is allowed).
When it is clear from the context we will drop the subscripts.
Let $R$ be the $\NP$ relation associated with the statement $\PreImage_{h,X_1,\ldots,X_{n+1}}$.

\subsection{Proving the Redacted Blockchain Consistency}
The above relation $R$ will now be used to show that all cases of deletion that we want to take into account can be reduced to proving and verifying in zero knowledge (ZK) the previous class of statements.

The first thing to notice is that if a transaction $\trans$  is redacted, the leaf in the Merkle tree representing the transaction $\trans$ subject to deletion would have an invalid hash (due to the fact that the illicit content has been replaced by zeros), so a node $P$ that wants to validate the block $B$ containing $\trans$ would reject $B$.
Here, it is where the NIZK proof $\pi$ comes into the play. Every time a full node $N$ has to delete data from a set of transactions $T=\{T_1,\ldots,T_l\}$ in a block $B$, $N$ executes the following steps.
\begin{enumerate}
	\item $N$ modifies the set of transactions $T$ generating a new set of transactions  $T'=\{T'_1,\ldots,T'_l\}$ where all data to be redacted in each transaction in $T$ are substituted  with zeroes.
	\item $N$ replaces $T$ with $T'$ in $B$.
	\item For each redacted transaction $T'_i$, $i\in\{1,\ldots,l\}$, in $T'$, $N$ generates a proof $\pi_i$ for the previous statement $T_i$. We remark that the replacement occurs only in {\em allowed} positions, that is in places where redaction is not harmful. Indeed the indexes of modified bytes are public, therefore anyone can check that these bytes are either data stored in an \verb|OP_RETURN| opcode or data stored in a \verb|scriptSig| of a coinbase transaction.
	\item $N$ deletes the redacted contents from her local Bitcoin blockchain.
	\item Every time someone requests $B$ to $N$, $N$ will send the blockchain containing $T'$ together with the generated proofs $\{\pi_1,\ldots,\pi_l\}$ and the statements for the proofs.
\end{enumerate}

By means of $\{\pi_1,\ldots,\pi_l\}$, $P$ can check that $B$, identified by a  \verb|Merkle Root|, is consistent with {\em some} set $T'$ of transactions that is identical to the set $T$ in $B$ except for some substrings (recall that the indices in which the substrings $y_i$'s occur and their lengths are public). $P$ runs $\Verify$ on input the public statement that depends only from $X_1,\ldots,X_{n+1}$, $h$ and the indices and the lengths of the deleted substrings (but does not need the actual deleted strings that are the witness known only to the prover). If the $\Verify$ procedure accepts the proof for each redacted transaction in each block, then $P$ can assume that the downloaded blockchain is consistent and can be used. Let us analyze the cases in which the redaction can be performed.

\paragraph{Deletion from input scripts, non-redeemable output scripts, and coinbase transactions}
 This can be the case of illicit content in coinbase transactions, illicit content of the type \verb|OP_RETURN <DATA>| nested inside a branch of an input script that is never executed and standard data output transactions that are not redeemable.

 In all such cases, the transaction has the form $s=X_1||y_1||\cdots X_n||y_n||X_{n+1}$ such that $H(s)=h$ and the substrings $y_1,\ldots,y_n$ represent illicit content.
Observe that the case $n>1$ models the possibility of having multiple  \verb|OP_RETURN <DATA>| occurrences nested inside a script or the fact that only some parts (but not all) of the string \verb|DATA| in a \verb|OP_RETURN <DATA>| or some parts in a \verb|scriptSig| of a coinbase transaction have to be deleted. 

Let $(\Prove,\Verify)$ be a NIZK for the previous relation $R$.
A node $A$ proceeds as follows. The node uses $\Prove$ to compute a NIZK proof $\pi$ for $R$ using the values $y_1,\ldots,y_n$ as witness and the known values $X_1,\ldots,X_{n+1}$ (along with the indices and lengths of the deleted strings) and then replaces all occurrences of $y_1,\ldots,y_n$ by zeros.
We remark that in case of deletion of data related to an \verb|OP_RETURN| opcode, the opcode itself and the length of bytes to store in the script will remain equal, the only change performed is on the bytes in \verb|<DATA>| that are substituted with 0's bytes. Moreover, notice that the \verb|scriptSig| of the coinbase transactions is not used to validate the transaction. In these cases the deleted data cannot belong to scripts that can be redeemed, therefore there is no risk that one of the modified transactions will cause a failure in the redeeming procedure.

\begin{figure}[htp] 
    \centering
    \subfloat[Merkle tree before deletion.]{
        \begin{tikzpicture}
[
    level 1/.style = {sibling distance = 4cm},
    level 2/.style = {sibling distance = 2cm},
    level 3/.style = {level distance = 1.2cm},
    every node/.style = {align=center,font=\footnotesize},
    edge from parent/.style={draw,latex-}
]
\node [draw,rounded corners]{Top Hash\\$H(H_0+H_1)$}
    child { node[draw,rounded corners] {$H_0$\\$H(H_{0,0}+H_{0,1})$}
    child { node[draw,rounded corners] {$H_{0,0}$\\$H(T_0)$}
    child { node {$T_0$}}}
    child {node[draw,rounded corners] {$H_{0,1}$\\$H(T_1)$}
    child { node {$T_1$}}} } 
    child { node[draw,rounded corners] {$H_1$\\$H(H_{1,0}+H_{1,1})$}
    child { node[draw,rounded corners] {$H_{1,0}$\\$H(T_2)$}
    child { node {$T_2$}}}
    child {node[draw,rounded corners] {$H_{1,1}$\\$H(T)$}
    child { node {$T$}}} } ;
 
\end{tikzpicture}
        \label{fig:a}
        }
    \hfill
    \subfloat[Merkle tree after deletion.]{
        \begin{tikzpicture}
[
    level 1/.style = {sibling distance = 4cm},
    level 2/.style = {sibling distance = 1.5cm},
    level 3/.style = {level distance = 1.2cm},
    every node/.style = {align=center,font=\footnotesize},
    edge from parent/.style={draw,latex-}
]
\node [draw,rounded corners]{Top Hash\\$H(H_0+H_1)$}
    child { node[draw,rounded corners] {$H_0$\\$H(H_{0,0}+H_{0,1})$}
    child { node[draw,rounded corners] {$H_{0,0}$\\$H(T_0)$}
    child { node {$T_0$}}}
    child {node[draw,rounded corners] {$H_{0,1}$\\$H(T_1)$}
    child { node {$T_1$}}} } 
    child { node[draw,rounded corners] {$H_1$\\$H(H_{1,0}+H_{1,1})$}
    child { node[draw,rounded corners] {$H_{1,0}$\\$H(T_2)$}
    child { node {$T_2$}}}
    child {node(AA)[draw,rounded corners] {$H_{1,1}$\\$H(T)$}
    child { node(BB)[red] {$T$}}
    child [anchor = west]{ node[draw,rounded corners,dashed] {$V(T',\pi)$}edge from parent[dashed]}}
    } ;
    \path (AA) -- (BB) 
    node[draw,cross out, pos=0.6] {};
    
\end{tikzpicture}
        \label{fig:b}
        }
    \caption{This figure represents the Merkle tree before and after that $T$ is modified. In (a) there is the Merkle tree before the deletion of data, in which all leaves are hashes and the root of the Merkle tree (the top hash) is equal to the root of the Merkle tree stored in the block of the blockchain. In (b) there is the modified transaction $T'$; if the hashes are not verifiable, then the verifier of the consistency of the block uses the proofs generated by our tool to check that $T'$ is a consistent modification of an harmless transaction.}
	\label{fig:merkle}
\end{figure}
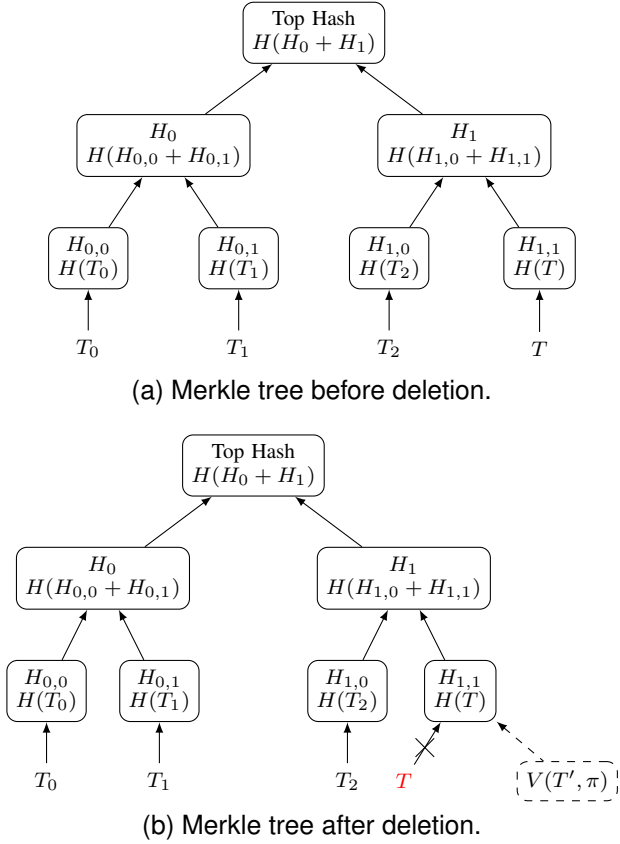

We illustrate what happens w.r.t. a Merkle tree in Figure~\ref{fig:merkle}, where $T$ is a transaction to redact, $T'$ is the redacted transaction, $\pi$ is the proof generated by our tool, and $V=(T',\pi)$ means that $T'$ is an input for the proof $\pi$ that produces a successful verification.

\paragraph{Deletion from output scripts that are redeemable}
In Bitcoin the \verb|Script| language has logical opcodes \verb|OP_IF|, \verb|OP_NOTIF|, \verb|OP_ELSE|, \verb|OP_END| that allow an \verb|OP_RETURN| to be set in a branch that is {\em never} executed.
In this case a redeemable output script \verb|out| can also contain a substring of the form \verb|OP_RETURN <DATA>|. A redeeming input script \verb|inp| of a subsequent transaction needs to sign a string $s$ that is the concatenation of \verb|out| with \verb|inp| in the redeeming transaction.
See for instance the script in Figure~\ref{fig:script}.
\begin{figure}
	\begin{center}
		\begin{boxedminipage}{\columnwidth}
			\begin{enumerate}
				\item \verb|OP_TRUE|
				\item \verb|OP_NOTIF|
				\item \verb|OP_RETURN <Illegal content>|
				\item \verb|OP_ENDIF|
				\item \verb|OP_DUP|
				\item \verb|OP_HASH160|
				\item \verb|<pubKeyHash>|
				\item \verb|OP_EQUALVERIFY|
				\item \verb|OP_CHECKSIG|
			\end{enumerate}
		\end{boxedminipage}
	\end{center}
	\caption{Script that contains a branch that is never executed in which arbitrary content can be stored.}
	\label{fig:script}
\end{figure}
The instruction $3$ is never executed whatever input script in a future transaction will try to redeem the above output script; only instructions $5$-$9$ will be executed (lines $5$-$9$ represent a standard way to send money from a user $A$ to a user $B$ in Bitcoin).

Observe that node $B$ has deleted the original string \verb|out| so is {\em unable} to verify the signature. Also other nodes that have downloaded the same block from $B$ do not have the redacted data, therefore are unable to validate the signature as well. 
To make our mechanism applicable in the case of redeemable transactions it is possible to tweak Bitcoin signature as being signature of the string $s$ as concatenation of \verb|H(out)| and \verb|H(inp)|.
Indeed, notice that whatever string $s$ has to be signed, the signing algorithm internally ``signs'' the digest \verb|H(s)|.
Therefore, we can tweak the \verb|OP_CHECKSIG| as follows. 
If \verb|out| is redacted then \verb|OP_CHECKSIG| checks if there is a NIZK proof 
$\pi$ of the fact that \verb|H(out)| is consistent with \verb|out| after the redaction (this statement can be expressed as a special case of the class of statements $\PreImage$ and thus $\Prove$ and $\Verify$ can be used as before for the same purposes). If \verb|H(out)| is correctly computed from \verb|out| and the proof of consistency is an accepting proof, \verb|OP_CHECKSIG| verifies that \verb|h=H(s)| is such that \verb|s| is the concatenation of \verb|H(out)| with \verb|H(inp)| and therefore uses the procedure described in Section~\ref{subsubsec:script}.
\subsection{Breaking the Generic Statement in more ``Efficient'' ad-hoc Statements}
We could implement our system using a STARK for the relation $R$ induced by the statement $\PreImage_{h,X_1,\ldots,X_{n+1}}$ described above. 
There are two problems with this approach, one theoretical and one practical.
The first problem is that, even if there is only a single and short deletion of few bytes, the time complexity of verifying the proof will depend on the length of the overall transaction and this is a wasteful overkill.
The second problem is that for larger transactions, the length of the corresponding rank-1 constraint system (R1CS), that is the constraint system used to represent a circuit,
becomes huge. 
The storage needed to store the R1CS for transactions of size greater than $1$KB, would consist of hundreds of gigabytes.

Instead of proving and verifying the previous statements directly in ZK (i.e., using a STARK for those statements), we prove and verify such statements in a more efficient way. 
The idea is to consider all intermediate outputs of each round of SHA256.
Recall that SHA256 essentially works as follows: given an input $X$, it extends $X$ to an input $X'$ of a length multiple of $64$ bytes, breaks $X'$ into chunks of $64$ bytes and for each of such chunks it executes a round function $\SHARound$ that takes as input a chunk and the output of the previous round.
The first round takes as input the first chunk and a fixed value $h_0$ that is the concatenation of values $g_0,\ldots,g_7$ described in the SHA256 specifications~\cite[Section 6]{shaspec}\footnote{In the specifications these variables are called $a$, $b$, $c$, $d$, $e$, $f$, $g$, $h$. We renamed these variables since we are indicating with letter $h$ the hash of the transaction to redact.}.

Let $X$ be a string obtained redacting a string $Y$ and let $h=H(Y)$. Recall that $X$ and $h$ are public information as well as the points in which the redaction has been done. The witness is the original string $Y$ before the redaction. Our goal is to design an efficient proof system to convince anyone that the public inputs are consistent with the redaction.

Let us say that SHA256 extends $Y$ (resp. $X$) into a string $Y'$ (resp. $X'$) consisting of $m$ chunks $Y_1,\ldots,Y_{m}$ (resp. $X_1,\ldots,X_{m}$) of $64$ bytes. 
The string $Y'$ (resp. $X'$) is obtained adding some bits at the end of the string $Y$ (resp. $X$) as prescribed by the SHA256 specifications~\cite[Section 5.1.1]{shaspec}. 
The prover will reveal the obtained intermediate outputs $h_1,\ldots,h_{m}=h$ of each round\footnote{Notice that it is not necessary for the prover to send all intermediate outputs, but only the ones corresponding to modified chunks, since the outputs of the unmodified chunks can be independently computed by the verifier.}, where for each $1\le i\le m$, $h_i=\SHARound(h_{i-1},Y_i)$.

Then, {\em only for one of the intervals subject to redaction}, the prover proves using a STARK that there exists a string $X_i$ such that $Y_i$ is the result of applying the redaction to $X_i$  and $h_i=\SHARound(h_{i-1},Y_i)$.
The verifier verifies each STARK relative to the chunks subject to redaction and for each other chunk $i$ not subject to redaction additionally verifies that $h_i=\SHARound(h_{i-1},Y_i)$; notice that the latter is verified just by running the round function on the known preimage.

\subsection{Security Analysis}
We can see the above STARK as a proof system for the class of statements $\PreImage_{h,X_1,\ldots,X_{n+1}}$ described above. Indeed, completeness and soundness are easy to check (see Section~\ref{sec:zk}).

We now analyze the security properties of our sanitizer tool. 
Consider a node $N$ who performs a deletion of some data from a transaction $t$ in a block $B$ stored on the blockchain.
Precisely, let $t=X_1||y_1||\cdots X_n||y_n||X_{n+1}$ such that $H(t)=h$, where the substrings $y_1,\ldots,y_n$ represent the illicit content and let $t'$ the resulting transaction after redaction. Notice that the redacted transaction $t'$ can be represented by just the values $X_1,\ldots,X_{n+1}$ (as before, for simplicity we omit the indices and length of the strings in which the deletion occurred).
$N$ will also add to the block a proof $\pi$ of the fact that the statement $\PreImage_{h,X_1,\ldots,X_{n+1}}$ holds. 

By the security of the hash function, a malicious node cannot deviate from the honest node by inserting a block $B'$ such that $B'$ and $B$ differ in places different from the redacted transaction.
Indeed, the verifier verifies the consistency of the blockchain from the genesis block until the block in which the redaction occurred and the consistency of the Merkle tree from the root to the redacted transaction.
Furthermore, by the soundness and the extractability of the STARK, if the proof $\pi$ is accepted by the verifier then the statement $\PreImage_{h,X_1,\ldots,X_{n+1}}$ is valid. This means that there exist substrings (known to the prover) $y_1,\ldots,y_n$ representing (possibly) illicit content and, by definition of the statement, the replacement occurred only in allowed places.
Therefore, the node can only redact content in allowed places and cannot compromise the UTXO database.
A similar argument works for multiple redactions.

Notice that revealing the intermediate outputs of the SHA256 function reveals whether two transactions have a common prefix. However, we adopt a pragmatic approach and do not consider harmful such  minor leakage.
We stress that for efficiency reasons, we do not compute proofs for the intervals not subject to the redaction.

We remark that our deletion technique does not require any joint decision to accept a redaction, therefore a redaction will not cause  any fork in the blockchain. Honest nodes will accept a given transaction independently from the fact that a subset of nodes individually and locally deleted some harmless content.

Moreover, our sanitizer tool makes unfeasible for an adversarial node to redact parts of the transactions that alter the UTXO database of Bitcoin. Indeed, it is always possible to check that the bytes from the starting position to the end position of each redeeming operation correspond either to the bytes of the \verb|<DATA>| field of an \verb|OP_RETURN| opcode or to the \verb|scriptSig| of a coinbase transaction. Therefore our redaction mechanism cannot be exploited to perform double-spending attacks.

\subsection{Multiple Deletions at Different Times} 
Breaking the statement to prove in more statements that exploit the round function \verb|SHARound| makes our solution more modular. Indeed in this case the prover will prepare a proof for each modified chunk in the SHA256 procedure instead of proving the knowledge of the preimage of the hash of the redacted transaction. The consequence of proving the knowledge of the preimage of \verb|SHARound| is that our  solution allows deletions of illicit content from different chunks of the same transaction at different times.
That is, a transaction $T_1$ in a block $B$ can be subject to deletion at time $t_1$ in a chunk $c_1$ and later at time $t_2$ the same transaction $T_1$ can be subject to deletion as well in a different chunk $c_2$. Moreover, our solution supports also the following scenario: a transaction $T_1$ in a block $B$ is redacted  at time $t_1$ and next at time $t_2$ a different transaction $T_2$ in the same block can be subject to redaction as well.

In our solution we do not consider the case in which the same chunk $c_1$ in a transaction $T_1$ has to be modified more than once since in this case it is not clear how to maintain efficiency.

\section{Our Implementation}\label{sec:implementation_eff}

In this section we introduce Isekai and then we illustrate our implementation of our Bitcoin sanitizer that uses Isekai. 

\subsection{Isekai}\label{def:isekai}
Our implementation is based on Isekai, a versatile framework for verifiable computation. Isekai allows to transform a C/C++ program into a set of R1CS constraints, an internal representation for many SNARKs/STARKs.
Moreover, Isekai offers an interface to several SNARK/STARK systems like the SNARK of~\cite{EC:GGPR13}, Bulletproof~\cite{SP:BBBPWM18} and Aurora~\cite{TCC:BenChiSpo16} allowing to invoke the prover and the verifier of such system in a black-box way.
\paragraph{Usage}
Isekai can generate a proof of the execution of a C/C++ function. The C/C++ function must have one of the following prototypes:

\begin{Verbatim}[breaklines=true]
void outsource(struct Input *input, struct NzikInput *nzik, struct Output *output);
void outsource(struct Input *input, struct Output *output);
void outsource(struct NzikInput *nzik, struct Output *output);
\end{Verbatim}

The variables \verb|input| and \verb|output| are public parameters and the variable  \verb|nzik| is the private input. 
The inputs are provided in an external file with the same name of the C/C++ program but with extension \verb|.in|. 
With the option \verb|--r1cs| the R1CS files are generated from the \verb|.in| file and then with these R1CS files it is possible to generate the proof using the \verb|--prove| option.
The proof is verified using the \verb|--verif| option. 
The specific SNARK/STARK scheme is chosen using the option \verb|--scheme|.

\subsection{Implementation of Our Bitcoin Sanitizer}
Here, we demonstrate the feasibility and practicality of our approach by providing a {\em sanitizer tool} that can be integrated in Bitcoin (or even other blockchains).
Our implementation shows another
application of ZK proofs that can be efficient enough to be used in practice. 

The goal of our tool is to show that our solution allows to perform redactions in {\em minutes} rather than {\em days} as in previous solutions.
We use STARKs combined with Isekai (see~\ref{def:isekai})  to convert C/C++ code into ZK proofs.
Among the available options, we selected  Aurora for the ZK proofs because Aurora provides:
a) Post-quantum security: Aurora is plausibly post-quantum secure (there are no known efficient quantum attacks against this construction) guaranteeing security even against future advances in quantum technology.
b) Fast verification: Aurora does not just provide short proofs but allows a verifier to run just in {\em logarithmic} time. 
c) Transparency: Aurora is transparent meaning that there is no trusted setup\footnote{Having a trusted setup in the context of Bitcoin would be questionable since Bitcoin should work without the need of any trusted party.}.
d) General C/C++ code: 
a publicly available C/C++ library of Aurora that supports R1CS is available\footnote{See \url{https://github.com/scipr-lab/libiop} (accessed 2022/10/04).} and this library is integrated into Isekai.
Our implementation is deployed for the Linux OS. 

We describe now the statement proved through a ZK proof by our implementation.
Let 
$X$ be the original transaction padded to a multiple of 64 bytes as described by SHA256 specifications~\cite{shaspec}; let $y_1,\ldots,y_m$ be the bytes to delete in $X$; let $Y$ be the transaction obtained substituting $y_1,\ldots,y_m$ in $X$ with bytes consisting of zeroes only, and padded as described by the SHA256 specifications~\cite{shaspec};
let	$\inte$ be the set of intervals in which $y_1,\ldots,y_m$ are modified in $X$; let $\SHARound$ be a circuit 
taking as input 1) a chunk of X, 2) the public values $g_0,\ldots,g_7$ described by SHA256 specifications~\cite[Section 6]{shaspec}~\footnote{In the specifications these variables are called $a$, $b$, $c$, $d$, $e$, $f$, $g$, $h$.}, 3) the output of the previous round.  $\SHARound$ produces new values $g_0',\ldots,g_7'$ as described by the SHA256 specifications.

We assume that $X=X_1,\ldots,X_n$ meaning that $X$ is composed by $n$ chunks of 64 bytes. The same holds for $Y$. Moreover, for simplicity, we define a function $f$ that given $Y$, $\inte$, and $y_1,\ldots,y_m$ is able to reconstruct the original $X$. The statement $\ChunkPreImage$ that our implementation proves for each modified block is the following: $\exists y_1,\ldots,y_m$ s.t. $\SHARound(g_0,\ldots,g_7, f(Y_i, \inte, y_1,\ldots,y_m))=g_0',\ldots,g_7'$, where $Y_i$, for $i\in\{1,\ldots, n\}$ is the modified chunk, the elements $y_1,\ldots,y_m$ form the witness owned by the prover and $g_0,\ldots,g_7, Y_i, \inte,g_0',\ldots,g_7'$ are all public values. We remark that the verifier can compute the output of all SHA256 rounds on unmodified blocks and check that the final hash is equal to the value stored in the Merkle tree of the block of the Bitcoin blockchain.

We now explain the content of our 
implementation describing how it works for a modified chunk $X_i$ of $X$.
The main function is:
\begin{Verbatim}[breaklines=true]
void outsource(struct Input *input, struct NzikInput *nzik, struct Output *output)
\end{Verbatim} 
that specifies: the public input \verb|input| of type \verb|struct Input *|, corresponding to the variables $X_i,g_0,\ldots,g_7$; the secret input \verb|nzik| of type \verb|struct NzikInput *|, corresponding to the variables $y_1,\ldots,y_n$; the output \verb|output| of type \verb|struct Output *|, corresponding to the variables $g_0',\ldots,g_7'$. The routine \verb|outsource| will use the public and private inputs to compute the intermediate hash of SHA256 on the current chunk and store it in \verb|output|\footnote{The code of the circuit corresponds to the one of SHA256.}.
The header file \verb|hash.h| specifies the types of the structures \verb|struct Input|, \verb|struct NzikInput| and \verb|struct Output|.
The structure \verb|struct Input| has the following format:
\begin{Verbatim}[breaklines=true]
struct Input {
	unsigned char trans[64];
	unsigned int g0[2];
	unsigned int g1[2];
	unsigned int g2[2];
	unsigned int g3[2];
	unsigned int g4[2];
	unsigned int g5[2];
	unsigned int g6[2];
	unsigned int g7[2];
	unsigned int start[64];
	unsigned int end[64]; };
\end{Verbatim}
while \verb|NzikInput| has the following format:
\begin{Verbatim}[breaklines=true]
struct NzikInput{
    unsigned char deleted_data[DEL_DATA_LENGTH]; }.
\end{Verbatim}

The field \verb|trans| contains the 64 bytes of $Y_i$, \verb|deleted_data| contains $y_1,\ldots,y_m$, and \verb|g0|, ..., \verb|g7| contain the output of the previous round of SHA256\footnote{If the modified chunk is the first chunk we note that \texttt{g0}, ..., \texttt{g7} are known and defined by the SHA256 specifications ~\cite[Section 5.3.3]{shaspec}.}. Isekai and Aurora work representing circuits so we have to fix an upper-bound to the maximum number of bytes that can be removed by a single 64 bytes chunk, that in our implementation is represented by the constant \verb|DEL_DATA_LENGTH| in the file \verb|hash.h|. The arrays \verb|start| and \verb|end| represent the starting points and the end points of each interval in which the data are removed.

From $Y_i$, the routine \verb|outsource| will first perform the string replacement using \verb|deleted_data|, \verb|start|, and \verb|end| obtaining back $X_i$. $X_i$ together with \verb|g0|, ..., \verb|g7| will be passed to $\SHARound$ to obtain the new values $g_0',\ldots,g_7'$ that will be put in \verb|struct Output| that is:
\begin{Verbatim}[breaklines=true]
struct Output { 
      unsigned int h_out[8]; };
\end{Verbatim}

If the number of deletion intervals is less than \verb|DEL_DATA_LENGTH|, the remaining elements of the arrays \verb|start| and \verb|end| can be set to $0$.
The program \verb|proofdel| contains all the routines used to prepare the data for Isekai.
The needed inputs for the prover are file \verb|original_tx|, file \verb|transaction|, and the intervals in which the user deleted the data.
File \verb|original_tx| is the file that contains the original transaction $T$ before the deletion.
File \verb|transaction| contains the transaction $T$ in which the bytes corresponding to the intervals taken in input by \verb|proofdel| are set to the byte $0X00$. 
On the other side, the needed inputs to verify the proofs, are the file \verb|transaction|, and the intervals in which the user deleted the data.
\verb|proofdel| will use these inputs to interact with Isekai to generate the circuit for the proofs, the proofs and to launch the verifier on each proof.
\verb|proofdel| interacts with Isekai to generate the proofs
computing the following steps:
 \begin{enumerate}
     \item perform the padding of the binary string contained in \verb|transaction| as prescribed by the SHA256 specifications~\cite{shaspec} and divide the padded transaction in chunks $C_0,\ldots, C_n$ of 64 bytes;
     \item take the original values of the data to delete from \verb|original_tx| and the hash of the transaction (before deletion);
     \item infer the  chunks $\{C_j\}_{j\in\{n\}}$ of the transaction $T$ that contain modified data, using the intervals, and then  for each of these $C_j$  prepare the public input and the witness to send to Isekai;
     \item receive back from Isekai a proof $\pi_j$ for each modified chunk $C_j$. 
 \end{enumerate}
 
 \verb|proofdel| will prepare the data to send to Isekai to verify the proofs through the following steps:
 \begin{enumerate}
     \item take in input the modified transaction for Bitcoin blockchain;
     \item recover the public inputs $\{i_j\}_{j \in \{n\}}$ and the proofs $\{\pi_j\}_{j \in \{n\}}$, where values $i_j$ is the index to the $j$-th deleted chunk and $\pi_j$ is the proof for the $j$-th chunk;
     \item send to Isekai the pairs $(i_j, \pi_j)$ for each modified chunk (after collecting all inputs and proofs), to start the verification procedure;
     \item end with success only if the verifier called by Isekai\footnote{We instantiate Isekai with Aurora.} accepts all the proofs.
 \end{enumerate}
 
For simplicity, we are omitting the further step in the verification procedure. Indeed, \verb|proofdel| will extract the intermediate output of SHA256 for each modified chunk from the public inputs and will use these intermediate outputs to compute the hash $h$ of $T$. If $h$ is equal to the value stored on the blockchain\footnote{To compute this step the user has to pass an additional parameter to the tool, that is the hash of the original $T$.} the verification procedure \verb|proofdel| succeds. Moreover, \verb|proofdel| will also check that deleted data are not contained in harmful positions. Indeed, in Bitcoin it is possible to check if a transaction is a coinbase transaction. In the affirmative case, one can check if deletion occurs only in a \verb|scriptSig|. If a deletion occurs after an \verb|OP_RETURN| opcode, \verb|proofdel| can check that the number of deleted bytes corresponds to the number of bytes contained in the  \verb|OP_RETURN| parameter. To check that redactions do not happen in harmful positions \verb|proofdel| does not need to interact with Isekai.

\section{Performance.}\label{sec:performance}

The system used to test our implementation consists of a desktop computer running  Ubuntu as operating system with an architecture $\times86\_64$, 32 GB of RAM and an Intel(R) Core(TM) $i7-7820X$ CPU 
with clockspeed 3.60GHz.
To execute our tests, we instantiate Isekai with Aurora. We remark that
our experiments only focused on evaluating the practical feasibility of our implementation, and our goal is not to test the performance of Isekai/Aurora. 

First, we analyzed the performance of our prover and verifier considering the number of modified chunks in a transaction.
Our tests show that the computations of prover and verifier are nearly linear as expected. The verifier runs in about 3 seconds to verify the proof of a single chunk of SHA256 and for each additional chunk to verify the same amount of time is required. Notice that a node must run the verifier only at {\em bootstrap} time.

We remark that our tests have not been optimized and the code would be highly parallelizable. In particular, in a cluster with $m>1$ processors  the time of the prover and verifier could be reduced approximately by a factor $m$ since the most expensive computation consists of running the prover and verifier on  independent statements.

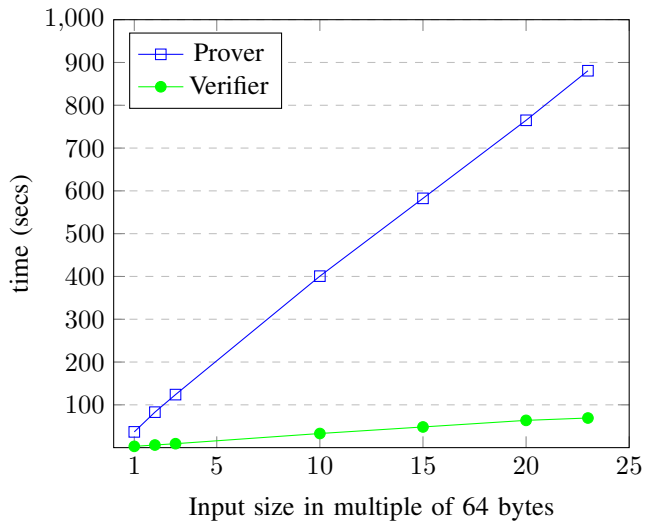
\begin{figure}
\begin{tikzpicture}
\begin{axis}[
	xlabel={Input size in multiple of 64 bytes},
	ylabel={time (secs)},
    xmin=0, xmax=25,
    ymin=0, ymax=1000,
    xtick={1,5,10,15,20,25},
	ytick={100,200,300,400,500,600,700,800,900,1000},
    legend pos=north west,
    ymajorgrids=true,
    grid style=dashed,
]

\addplot[
    color=blue,
    mark=square,
    ]
    coordinates {
    (1,36.8)(2,82.9)(3,123.9)(10,400.7)(15,582.5)(20,764.9)(23,880.7)
    };
\addplot[
    color=green,
    mark=*,
    ]
    coordinates {
    (1,3)(2,6.1)(3,9.2)(10,33.0)(15,48.3)(20,63.7)(23,69.2)
    };
\legend{Prover,Verifier}

\end{axis}
\end{tikzpicture}
\caption{Prover and verifier time on inputs of different size.}
\label{fig:prv_vrf_time}
\end{figure}
We tested our code on both real transactions taken from Bitcoin blockchain and on our own standard ad-hoc transactions. 

Our own transactions have
the purpose of evaluating our tool on different numbers of redacted chunks. Indeed for our tests we needed
data to delete in many consecutive and non-consecutive chunks, instead of restricting ourselves to what is available on the Bitcoin blockchain.

We now describe the transactions that we have considered in the performance evaluation. 
The 1st transaction is a 64 bytes transaction that we call ``Simple''. In Simple only 4 bytes contained in the first chunk were deleted. The 2nd transaction is the coinbase transaction of the genesis block. On this transaction we deleted the 69 bytes of the Chancellor sentence. We call this transaction ``Chanc''.
The 3rd transaction taken into account is Bitcoin transaction indexed ``$db27236623f19ceaf8535407e74b5dfad613aef7d555\allowbreak863 \allowbreak 1f4837fd0f6d83c83$'' in which we deleted 76 bytes distributed in 3 chunks. We call this transaction ``db2723''.
We define 4 ad-hoc \verb|OP_RETURN| transactions. We call them ``Ex1'', ``Ex2'', ``Ex3'' and ``Ex4'' respectively. The sizes of these transactions are respectively 1280, 1280, 2560 and 3888 bytes. We deleted 640 bytes from  Ex1 that were distributed in 10 SHA256 chunks, 920 bytes from Ex2 that were distributed in 15 SHA256 chunks, 1231 bytes from Ex3 that were distributed in 20 SHA256 chunks and 576 bytes Ex4, where bytes to delete  were distributed in 16 different \verb|OP_RETURN| output scripts contained in 23 SHA256 chunks.

The performance analysis reports the transaction length in bytes, the number of modified chunks, the number of bytes deleted by the entire transaction and the execution time in seconds of  prover and verifier. Results are shown in Table~\ref{tab:perf}. 

Specifying both the number of bytes redacted and the number of chunks allows to better scrutinize the performance of our tool.
Indeed, as expected, the execution time of the prover and the verifier grows linearly in the number of chunks modified in the transaction, and not with the total number of bytes redacted.
We note that even though in Ex3 there are 1231 deleted bytes, the time needed to generate the proofs is less than the time needed to generate the proofs for Ex4 where the number of deleted bytes is 576 bytes. This is caused by the number of deleted chunks, that are 20 for Ex3 and 23 for Ex4.
A graph of the execution time to generate the proofs and to verify them is reported in Figure~\ref{fig:prv_vrf_time}. 

\begin{table}
\caption{Performance of deletion in our tool. In this table we report the execution time of the prover and the verifier\label{tab:perf}}
\centering
\begin{tabular}{|c|c|c|c|c|c|}
	\hline
		      Tx name & Bytes & \shortstack{Modified\\ chunks \\(num.) }& \shortstack{Deleted \\bytes} & \shortstack{Prover \\(sec.)}&\shortstack{ Verifier \\(sec.)}\\
		       \hline
Simple     & 64                 & 1                      & 4                      & 36.8         & 3.0            \\
\hline
Chanc & 204                & 2                      & 69                     & 82.9         & 6.1            \\
\hline
db2723     & 283                & 3                      & 76                     & 123.9        & 9.2            \\
\hline
Ex1    & 1280              & 10                     & 640                    & 400.7        & 33.0            \\
\hline
Ex2    & 1280              & 15                     & 920                    & 582.5        & 48.3            \\
\hline
Ex3    & 2560              & 20                     & 1231                   & 764.9        & 63.7            \\
\hline
Ex4    & 3888              & 23                     & 576                    & 880.7        & 69.2            \\
\hline
\end{tabular}
\end{table}

\begin{figure}

\begin{tikzpicture}
\begin{axis}[
	xlabel={Transaction ID},
	ylabel={MB},
    xmin=0, xmax=6,
    ymin=12.1, ymax=12.500000,
    xtick={0,1,2,3,4,5,6},
	ytick={12.100000,12.200000,12.300000,12.4000000,12.500000},
    legend pos=north west,
    ymajorgrids=true,
    grid style=dashed,
]

\addplot[
    color=blue,
    mark=square,
    ]
    coordinates {
    (0,12.24)(1,12.25)(2,12.28)(3,12.29)(4,12.32)(5,12.34)(6,12.36)
    };
\addplot[
    color=green,
    mark=*,
    ]
    coordinates {
    (0,12.18)(1,12.23 )(2,12.26)(3,12.27)(4,12.30)(5,12.34)(6,12.43)
    };
\legend{Prover,Verifier}

\end{axis}
\end{tikzpicture}
\caption{Memory consumption for tested transactions.}
\label{fig:mem_cons}
\end{figure}
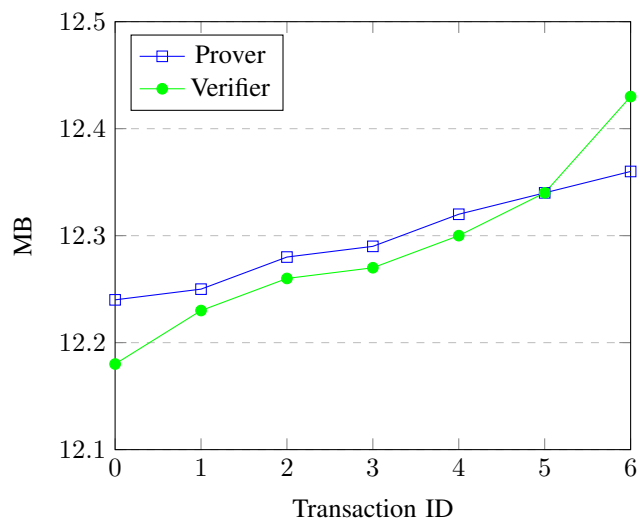

The memory usage of our tool  changes only slightly in the reported executions. Indeed, the prover of our tool repeats multiple times the generation of the proof sequentially on multiple redacted chunks and each of this generation requires the same amount of memory. Similarly, when considering the memory usage for the verification we note that our tool calls multiple times the verification procedure sequentially. Therefore the memory consumption of our tool is pretty much the same in each execution, and quite limited ($<13$ MB). See Table~\ref{tab:mem_cons_tab} and Figure~\ref{fig:mem_cons} for further details.

Notice that the memory usage is not the same as the storage used by a node who performs a redaction. Indeed, the additional storage required by a node with respect to the standard Bitcoin protocol will consist of the storage required to save the proofs generated by Aurora that are short ($< 130$ KB). Also, the additional required storage will be proportional to the number of redactions performed (one proof for each redaction).
\begin{table}
	\caption{Memory consumption in our tool. In this table we report the memory consumption of the prover and the verifier \label{tab:mem_cons_tab}}
\centering
\begin{tabular}{|c|c|c|c|}
\hline
Tx ID & Tx name  & Prover (MB) & Verifier  (MB) \\ \hline
0              & Simple     & 12.24                       & 12.18                         \\ \hline
1              & Chanc & 12.25                       & 12.23                         \\ \hline
2              & db2723     & 12.28                       & 12.26                         \\ \hline
3              & Ex1   & 12.29                       & 12.27                         \\ \hline
4              & Ex2   & 12.32                       & 12.30                         \\ \hline
5              & Ex3   & 12.34                       & 12.34                         \\ \hline
6              & Ex4   & 12.36                       & 12.43                         \\ \hline
\end{tabular}
\end{table}
The last data that we report describes the length of a single proof file when the number of modified bytes inside a single SHA256 chunk increases. We generated a binary string of 55 bytes for a single SHA256 chunk, and we modified all the bytes of the string starting from the first one and adding at each execution one more byte to delete\footnote{Our tool
can be used to prove knowledge of a SHA256 preimage of every binary string; we exploit this fact to analyze how the proof size variates 
depending on the number of bytes to modify.}. The smallest proof file generated consisted of 419 KB and was obtained when we deleted 32 bytes from the transaction, while the largest proof file size consisted of 431 KB and was obtained when we deleted 39 bytes from the transaction.

We remark that our solution scales with the number of total redactions performed by  a node.
For each redaction the complexity is a function of the number of modified chunks in the redacted transaction. Indeed, there is a proof for each modified chunk in each transaction. We stress that redactions are expected to be required only once in a while.

\begin{IEEEbiographynophoto}
{Vincenzo Botta}
is a research assistant in computer science at the University of Salerno, in  Fisciano, SA, 84084 Italy. Contact him at botta.vin@gmail.com.
\end{IEEEbiographynophoto}

\begin{IEEEbiographynophoto}
{Vincenzo Iovino}  
worked on this paper while he was an assistant professor at the University of Salerno, in Fisciano, SA, 84084 Italy. Contact him at viovino@unisa.it.
\end{IEEEbiographynophoto}
\begin{IEEEbiographynophoto}
{Ivan Visconti} is a full professor of Computer Science at the University of Salerno, in Fisciano, SA, 84084 Italy. Contact him at visconti@unisa.it. 
\end{IEEEbiographynophoto}
\end{document}